\documentclass[aps,prl,10pt,citeautoscript,twocolumn,showpacs,superscriptaddress]{revtex4-2} 
\usepackage{graphicx}
\usepackage{amsmath}
\usepackage{color}
\usepackage{float}
\usepackage{placeins}
\usepackage[normalem]{ulem}
\usepackage[retainorgcmds]{IEEEtrantools}
\usepackage{natbib}
\usepackage{xspace}
\usepackage{physics}
\usepackage[colorlinks,linktocpage,bookmarks=false]{hyperref} 
\usepackage[table,dvipsnames]{xcolor}

\newcommand\myshade{85}
\definecolor{myrulecolor}{RGB}{150,20,0}
\colorlet{mylinkcolor}{violet}
\colorlet{mycitecolor}{YellowOrange}
\colorlet{myurlcolor}{Aquamarine}
\hypersetup{
	linkcolor  = myurlcolor!\myshade!black,
	citecolor  = mycitecolor!\myshade!black,
	urlcolor   =  myrulecolor!\myshade!black,
}

\begin{document}

\newcommand{\rt}{R$_2$T$_2$O$_7$\xspace}
\newcommand{\lahf}{La$_2$Hf$_2$O$_7$\xspace}
\newcommand{\cesn}{Ce$_2$Sn$_2$O$_7$\xspace}
\newcommand{\cezr}{Ce$_2$Zr$_2$O$_7$\xspace}
\newcommand{\cehf}{Ce$_2$Hf$_2$O$_{7}$\xspace}
\newcommand{\przr}{Pr$_2$Zr$_2$O$_7$\xspace}
\newcommand{\prsn}{Pr$_2$Sn$_2$O$_7$\xspace}
\newcommand{\prhf}{Pr$_2$Hf$_2$O$_7$\xspace}
\newcommand{\ndzr}{Nd$_2$Zr$_2$O$_7$\xspace}
\newcommand{\hoti}{Ho$_2$Ti$_2$O$_7$\xspace}
\newcommand{\dyti}{Dy$_2$Ti$_2$O$_7$\xspace}
\newcommand{\tbti}{Tb$_2$Ti$_2$O$_7$\xspace}
\newcommand{\tbsn}{Tb$_2$Sn$_2$O$_7$\xspace}
\newcommand{\ce}{Ce$^{3+}$}
\newcommand{\cebis}{Ce$^{4+}$}
\newcommand{\pr}{Pr$^{3+}$}
\newcommand{\yb}{Yb$^{3+}$}
\newcommand{\degC}{$^\circ$C}
\newcommand{\ab}[1]{{\color{red} #1}}

\title{Dipolar-octupolar correlations and hierarchy of exchange interactions in Ce$_2$Hf$_2$O$_7$}
\author{Victor Por\'{e}e}
\email[]{victor.poree@psi.ch}
\affiliation{Laboratory for Neutron Scattering and Imaging, PSI Center for Neutron and Muon Sciences, Paul Scherrer Institut, 5232 Villigen PSI, Switzerland}
\author{Anish Bhardwaj}
\email[]{anis.bhardwaj@gmail.com}
\affiliation{Department of Physics, Florida State University, Tallahassee, FL 32306, USA}
\affiliation{National High Magnetic Field Laboratory, Tallahassee, FL 32310, USA}
\affiliation{Department of Physics, St. Bonaventure University, New York 14778, USA}
\author{Elsa Lhotel}
\affiliation{Institut N\'eel, CNRS, Universit\'e Grenoble Alpes, 38042 Grenoble, France}
\author{Sylvain Petit}
\affiliation{LLB, CEA, CNRS, Universit\'{e} Paris-Saclay, CEA Saclay, 91191 Gif-sur-Yvette, France}
\author{Nicolas~Gauthier}
\affiliation{Institut Quantique, D\'epartement de physique, and RQMP, Universit\'e de Sherbrooke, Sherbrooke, Qu\'ebec J1K 2R1, Canada}
\author{Han Yan}
\affiliation{Department of Physics \& Astronomy, Rice University, Houston, TX 77005, USA}
\affiliation{Smalley-Curl Institute, Rice University, Houston, TX 77005, USA}
\affiliation{Institute for Solid State Physics, The University of Tokyo. Kashiwa, Chiba 277-8581, Japan}
\author{Vladimir Pomjakushin}
\affiliation{Laboratory for Neutron Scattering and Imaging, PSI Center for Neutron and Muon Sciences, Paul Scherrer Institut, 5232 Villigen PSI, Switzerland}
\author{Jacques Ollivier}
\affiliation{Institut Laue-Langevin, CS 20156, F-38042 Grenoble Cedex 9, France}
\author{Jeffrey A. Quilliam}
\affiliation{Institut Quantique, D\'epartement de physique, and RQMP, Universit\'e de Sherbrooke, Sherbrooke, Qu\'ebec J1K 2R1, Canada}
\affiliation{Université Paris-Saclay, CNRS, Laboratoire de Physique des Solides, 91405, Orsay, France}
\author{Andriy H. Nevidomskyy}
\affiliation{Department of Physics \& Astronomy, Rice University, Houston, TX 77005, USA}
\author{Hitesh J. Changlani}
\email[]{hchanglani@fsu.edu}
\affiliation{Department of Physics, Florida State University, Tallahassee, FL 32306, USA}
\affiliation{National High Magnetic Field Laboratory, Tallahassee, FL 32310, USA}
\author{Romain Sibille}
\email[]{romain.sibille@psi.ch}
\affiliation{Laboratory for Neutron Scattering and Imaging, PSI Center for Neutron and Muon Sciences, Paul Scherrer Institut, 5232 Villigen PSI, Switzerland}

\begin{abstract}
\noindent 
High-resolution neutron spectroscopy on \cehf reveals a correlated state characterized by distinct dipolar scattering signals -- quasi-elastic and inelastic contributions consistent with `photon' and `spinon' excitations in quantum spin ice. These signals coexist with weak octupolar scattering.
Fits of thermodynamic data using numerical methods indicate a dominant octupolar exchange, $J_{x}$ or $J_{y}$, with substantial dipolar $J_{z}$ and minute dipole-octupole $J_{xz}$ couplings. The $J_{xz}$ value is corroborated by an independent fit of the neutron scattering amplitude balance between dipolar and octupolar `photon' contributions, highlighting its importance to understand neutron scattering results in this family.
\cehf enriches the landscape of dipole-octupole pyrochlore physics, and reveals a `quantum multipolar liquid' where hybrid correlations involve multiple terms in moment series expansion, opening questions on their intertwining and hierarchy in quantum phases.

\end{abstract}
\maketitle

% Introduction ==========================================================

Multipoles in condensed matter refer to higher-order terms of a series expansion describing distributions of electric and magnetic charges. They originate from local arrangements of magnetic dipoles (`cluster multipoles')~\cite{Suzuki17,Huebsch21,Kimata2021,Bhowal22} or elements with unquenched orbital moments~\cite{Santini09,Kuramoto09}. Strong spin-orbit coupling can result in multipolar phases in compounds of \textit{f} elements~\cite{Mydosh2020,Shiina97,Santini06} or heavy \textit{d} elements~\cite{Chen10,Hirai20}, although experimental verification is difficult as multipoles tend to remain hidden for conventional scattering techniques.
Long-range magnetic structures involving these elements can lead several of the symmetry-allowed multipoles to order, with debates on the nature and hierarchy of order parameters~\cite{Iwahara22,PourovskiiPNAS2021}. Liquid correlations involving magnetic dipoles and electric quadrupoles were argued in Pr$_2$Zr$_2$O$_7$ based on bulk properties~\cite{Tang2023} and inelastic neutron scattering in samples with residual disorder~\cite{PZO_Sylvain,PhysRevX.7.041028}.
However, directly probing correlations of different multipoles in a liquid phase remains elusive.
\\

\begin{figure*}
\includegraphics[width=15 cm]{Fig1_New_v2.png}
\centering
\caption{(a) INS spectra measured between 0.1~K and 5~K and integrated between 0.2 and 1.0~\AA$^{-1}$. (b)~$\chi^{\prime\prime}(E)$ (points with error bars), obtained by subtracting 5~K data. Red lines are fits using a Lorentzian peak shape for comparison with previous data on \cezr~\cite{CZO_US,CZO_CA} and \cesn~\cite{CSO_NatPhy}. (c)~Difference map of the magnetic dynamical structure factor $S(Q,E)$ between 0.1~K and 5~K. (d)~and (e) are constant energy cuts at 0 $\pm$ 0.01~meV and 0.03 $\pm$ 0.01~meV energy transfers, respectively. Black and red lines on panel d are fits using model calculations of ``quasi-elastic'' scattering for classical and quantum spin ice, respectively. (f)~High-$Q$ diffuse scattering obtained from the difference between neutron diffraction patterns measured at set-point temperatures of 5~K and 0.05~K. Intensities are normalized to nuclear scattering as detailed in~\cite{CSO_NatPhy}. The red line is based on theoretical calculations assuming ice rules among mixed dipolar-octupolar wavefunctions (see~\cite{supp})}.
\label{Fig1} 
\end{figure*}

In compounds of $f$ elements, the number of exchange parameters between spin-orbital entangled $J$ multiplets can be large. However, in rare-earth insulators such as pyrochlore oxides -- frustrated magnet prototypes for the quantum spin ice (QSI) state~\cite{Hermele:2004gg,Onoda2010,Savary:2012cq,Shannon2012,Benton:2012ep,Gingras:2014ip,pace2021,laumann2023}, interactions occur on energy scales that are small enough to involve only the ground state doublet~\cite{Rau19}. 
The local $D_{3d}$ symmetry in pyrochlores gives rise to three possible kinds of ground-state doublets, depending on the number of \textit{f} electrons and crystal-electric field (CEF). 
One such possibility is the `dipole-octupole' doublet~\cite{DO-PRL,Rau19}, stabilized e.g. in Ce$^{3+}$ pyrochlores~\cite{DO-PRB,PhysRevResearch.5.033169}, in which case it is defined by any linear combination of $|m_J=\pm3/2\rangle$ states.
Projection onto the subspace spanned by the two elements of this doublet defines a pseudo-spin 1/2 with $(s^x, s^y, s^z)$ components \cite{DO-PRL,DO-PRB,supp,PhysRevResearch.5.033169}. Due to the specific transformations of this doublet under the local symmetries, a minimal model of interactions is given by the XYZ Hamiltonian
%%%%%%%%%%%%%%%%%%%%%%%%%%%%%%%%%%%%%%%%%%%%%%%%%%%%%%%%%%%%
\begin{eqnarray}\label{eq:ham}
H_{nn}&=&\sum_{\langle ij \rangle} J_{x} s_i^x s_j^x+J_{y}s_i^ys_j^y+J_{z}s_i^zs_j^z +J_{xz}(s_i^xs_j^z+s_i^zs_j^x)   \nonumber \\
&&-\underset{i}{\sum}\left(\hat{z}_i\cdot {\bf{h}}
\right) g_{z}s_i^z
\end{eqnarray}
where $J_{x},\ J_{y},\ J_{z}$ and $J_{xz}$ are effective coupling strengths, and the summation $\langle ij \rangle$ is over nearest-neighbors \cite{DO-PRL}. 
%%%%%%%%%%%%%%%%%%%%%%%%%%%%%%%%%%%%%%%%%%%%%%%%%%%%%%%%%%%%
$s^z$ is reflected in the magnetic moment and in the dipolar neutron cross-section, while $s^x$ and $s^y$ are reflected in the orbital neutron cross-section and appear at large momentum transfer $Q$. QSI phases emerge when $J_{x}$, $J_{y}$ or $J_{z}$ is ferromagnetic and dominant, leading to a manifold of \textit{ice} states, while other terms bring quantum fluctuations~\cite{DO-PRL} with the possibility of achieving tunable emergent quantum electrodynamics \cite{sanders202arXiv}.

Cerium pyrochlores are the focus of intense investigations for the study of dipolar-octupolar (DO) QSI. \cesn~was first studied and samples obtained by solid-state synthesis show a correlated phase below 1~K based on thermodynamic measurements, while muon spin relaxation excludes long-range magnetic order down to 0.02~K~\cite{CSO_PRL}.
Neutron scattering experiments reveal a continuum of excitations composed of three bands as expected for the $\pi$-flux phase of QSI~\cite{CSO_spinons,PhysRevLett.132.066502}, and a strengthening of the octupolar moment explained by a dominant $J_{y}$ or $J_{x}$ \cite{CSO_NatPhy}.
However, other studies using samples prepared hydrothermally show different scattering, reminiscent of a spin ice regime of magnetic dipoles that would be a proximate state to long-range antiferromagnetic order~\cite{CSO_CA}. In \cezr, neutron scattering and single crystal specific heat investigations concluded to a QSI, with $J_x\sim J_y$ ~\cite{CZO_CA,CZO_US,CZO_CA_2,CZO_US_2,CZO_US_3}. Recent attempts were made to discern the quasi-elastic `photon' and inelastic `spinon' scatterings~\cite{CZO_US_4}.
A comparison of the low-temperature specific heat in \cesn~\cite{CSO_NatPhy}, \cezr~\cite{CZO_US} and \cehf~\cite{CHO_PRM} exemplifies intrinsic differences. The relative strength of the exchange interactions and their effects on experimental observations, as well as sample dependencies, are central questions in this remarkable series of materials.

Here we investigate the correlated state in \cehf~\cite{CHO_PRM}, revealing signatures of dipolar-octupolar correlations.
Consistently, the determination of the exchange couplings from fits of thermodynamic data points to a DO-QSI ground state. In addition, high resolution inelastic neutron scattering measurements highlight a clear separation between the expected photon bandwidth and the spinon excitations, and confirm the parameters of the Hamiltonian.
 
\begin{figure*}[!ht]
\centering
\includegraphics[scale=0.335]{Fig_2_new_CpScaled.png}
\centering
    \caption{(a-c) Magnetic contribution to the specific heat, (d) magnetization and (e)~effective magnetic moment derived from magnetic susceptibility, with magnetic fields applied along $[111]$. All dots are experimental data and curves are results of the Lanczos analysis (16-sites system) for two of the best sets of interactions. The fits $a$, $b$ and $c$ were obtained with $g_{z}$ = 2.328 and ($J_{x}$,~$J_{y}$,~$J_{z}$,~$J_{xz}$)$_{a}$~=~(0.011,~0.044,~0.016,~-0.002), ($J_{x}$,~$J_{y}$,~$J_{z}$,~$J_{xz}$)$_{b}$~=~(0.020,~0.047,~0.013,~-0.008) or ($J_{x}$,~$J_{y}$,~$J_{z}$,~$J_{xz}$)$_{c}$~=~(0.046,~0.022,~0.011,~-0.005) in unit of meV. (f) The log of the two dimensional $J_{z}$-$J_{x}$ cost function obtained by fixing  $J_{y}=0.047$ meV, $J_{xz}=-0.008$ meV, and $g_{z}=2.328$. The dashed curve encircles solutions where the log of the cost function is less than -0.42, highlighting the best solutions.}
\label{Fig_Cp}
\end{figure*}
 
% Methods =============================================================
Specific heat data were measured using a Quantum Design PPMS, in zero and finite magnetic fields up to 6~T applied along the crystallographic $[111]$ direction, in a temperature range from 0.4 to 15~K. The lattice contribution was subtracted using data measured for \lahf. Additional data were taken between 0.05~K and 0.8~K using a home-built calorimeter in a dilution refrigerator and the quasi-adiabatic heat pulse method. The heater and thermometer were fixed directly to the sample and contacts were made with 7~$\mu$m diameter NbTi wires to minimise heat leaks.
Magnetization \textit{vs} field was measured using SQUID magnetometers equipped with a miniature dilution refrigerator developed at the Institut N\'eel-CNRS Grenoble~\cite{Paulsen01}. Neutron powder diffraction was performed on HRPT (SINQ)~\cite{HRPT} using a wavelength of 1.15~\AA~and a powder sample in a dilution refrigerator. 
Inelastic neutron scattering (INS) data were collected using a powder sample on IN5 (ILL) using an incident energy of 0.82~meV, providing a resolution of 11 $\mu$eV. All measurements used samples reported in Ref.~\onlinecite{CHO_PRM}.

% ResultsAndDiscussion ==========================================================
% INS
INS data integrated over low $Q$ values reveal the presence of an inelastic signal (Fig.~\ref{Fig1}(a)), similar to the continua of spinon excitations in \cesn~\cite{CSO_NatPhy,CSO_spinons} and \cezr~\cite{CZO_US,CZO_US_2,CZO_US_4}. Spectra were collected at temperatures inside and outside the correlated regime. 
The high temperature spectrum was subtracted to extract the imaginary part of the generalized dynamic spin susceptibility $\chi^{\prime\prime}(E)=[1-\exp(-E/{k_{B}T})]S(E)$, with $S(E)$ the magnetic dynamical structure factor (Fig.~\ref{Fig1}(b)). 
The signal rapidly decreases upon warming from 0.1~K to 0.2~K, and has almost vanished at 0.4~K, consistent with the weak energy scale in cerium pyrochlores. 
The band of excitations is centered around $\Delta$~=~0.024~$\pm$~0.002~meV -- a significantly smaller energy compared to $\Delta$~=~0.039~$\pm$~0.003~meV in \cesn~\cite{CSO_NatPhy} and $\Delta$~$\sim$~0.04~meV in \cezr~\cite{CZO_US}. 
The width of the inelastic signal ($\Gamma$~$\sim$~0.01~meV) is sharper than in both \cesn~($\Gamma$~$\sim$~0.025~meV) and \cezr~($\Gamma$~$\sim$~0.06~meV).
For spinon excitations in a QSI -- magnetic monopoles endowed with quantum dynamics~\cite{Huang18,Udagawa19,Morampudi20,DO-PRB,PhysRevResearch.5.033169,PhysRevB.96.085136,PyrochloreU1GangChen,Hosoi22,Desrochers22,PhysRevLett.132.066502}, the center of the band is set by the energy scale of the dominant interaction while its width relates to transverse couplings responsible for quantum fluctuations. 
However, conclusions made on the basis of the values of $\Delta$ and $\Gamma$~consider a generic QSI spin-1/2 quantum XYZ model~\cite{Onoda11} with dominant and transverse exchanges, which may not reflect all subtleties of the four exchange parameters in Hamiltonian \eqref{eq:ham}~\cite{DO-PRL}.\

% HRPT
Having identified excitations that possibly indicate a QSI phase, a legitimate question arising for a `dipole-octupole' pyrochlore is the nature of the underlying correlations, i.e. of the dominant coupling. We have performed a thermal neutron powder diffraction experiment in the same conditions as for \cesn~\cite{CSO_NatPhy}, to search for octupolar correlations. The result is shown in Fig.~\ref{Fig1}(f) together with the data previously reported for \cesn~\cite{CSO_NatPhy}. Using the same procedure for scaling in absolute units as in Ref.~\cite{CSO_NatPhy}, we found that the same type of high-$Q$ scattering occurs in \cehf~at low temperature, but with a much weaker intensity than in \cesn. 
% Back to IN5
However, looking at the INS data (Fig.~\ref{Fig1}(c-d)), we also observe at low-$Q$ a signal centered at $E=0$, within the instrumental resolution $\pm 11 \mu$eV, which we call hereafter `quasi-elastic', and suggesting the simultaneous existence of dipolar correlations. This contrasts with INS experiments performed on \cesn prepared by solid-state synthesis, where no quasi-elastic dipolar signal is observed and the scattering at low $Q$ only comprises gapped spinon excitations~\cite{CSO_NatPhy,CSO_spinons}.

To rationalize the neutron scattering signals, we then estimate $J$ parameters using fits of specific heat, magnetization and susceptibility. 
The magnetic specific heat was fitted using the finite temperature Lanczos method (FTLM) \cite{FTLM.Prelovsek,Changlani.YTO} applied to the dipole-octupole QSI Hamiltonian in Eq.~\eqref{eq:ham}~(see Ref.~\cite{supp,CZO_US_2}).
The $g$-tensor components were initially estimated from magnetization \textit{vs} field data collected at 4~K, far from the correlated regime. The resulting $g_{z}$~=~2.328 agrees with expectations from the CEF~\cite{supp} and was fixed during the optimization of $J$ values.
The zero-field specific heat (Fig.~\ref{Fig_Cp}(a)) shows a broad peak, centered around 0.15~K, as typically observed in spin ice materials~\cite{Matsuhira2009,Kimura:2013gj,PZO_Sylvain,PHO_PRB,CSO_NatPhy,CZO_US}.
Applying a magnetic field along the [111] direction shifts the signal to higher temperatures, where two separate contributions develop, whose positions and relative weights are remarkably captured by the simulations \cite{supp}. 
Calculations from our model were also compared with the temperature dependence of the bulk susceptibility measured at low field and magnetization curves measured at 0.08~K for fields along the high-symmetry directions~\cite{supp}, giving a relatively good agreement. The former is shown in Fig.~\ref{Fig_Cp}(e) using a highly discriminating plot -- the effective magnetic moment \textit{vs} temperature on a logarithmic scale, showing that the drop of dipole moment in the correlated regime is reproduced for dominant octupolar couplings.
The cost function resulting from our analysis~\cite{supp} is presented in Fig.~\ref{Fig_Cp}(f) for a dominant $J_{y}$, showing a valley of optimal parameter sets that correspond to an octupolar QSI. Two representative sets (labeled $a$ and $b$ and respectively shown as red and violet curves in Fig.~\ref{Fig_Cp}(a-e)) are (all $J$ values in meV):
\begin{equation}
\label{eq:optimal-params}
    \begin{split}
        (J_{x},J_{y},J_{z},J_{xz})_{a} = (0.011,0.044,0.016,-0.002),\\
        (J_{x},J_{y},J_{z},J_{xz})_{b} = (0.020,0.047,0.013,-0.008).
    \end{split}
\end{equation} 
Importantly, dominant $J_{x}$~solutions are also valid, e.g.:
 \begin{equation}
\label{eq:optimal-params2}
    (J_{x},J_{y},J_{z},J_{xz})_{c}=(0.046,0.022,0.011,-0.001).
\end{equation}
%%%%%%%%%%%%%%%%%%%%%%%%%%%%%%%%%%%%%%%%%%%%%%%%%%%%%%%%%%%
These parameters are consistent with our analysis from data measured with magnetic fields along [110]~\cite{CHO_Bhardwaj}. Close $J$ values are also reported from numerical linked cluster calculations obtained from zero-field specific heat of another crystal~\cite{CHO_Smith}. In all our optimal parameter sets, either $J_x$ or $J_y$ dominates, which contrasts with \cezr where $J_{y}\approx J_{x}$~\cite{CZO_CA_2,CZO_US_2}.
%%%%%%%%%%%%%%%%%%%%%%%%%%%%%%%%%%%%%%%%%%%%%%%%%%%%%%%%%%%
Nonetheless, our parameters correspond to the $\pi$-flux phase of QSI~\cite{Lee2012,PhysRevB.96.085136,PyrochloreU1GangChen,PhysRevLett.121.067201}, like in \cezr~\cite{CZO_CA_2,CZO_US_2} and solid-state synthesized \cesn~\cite{CSO_NatPhy,CSO_spinons}: transverse interactions lead to $J_{\pm} < 0$, calculated as $-(J_{x}+J_{z})/4$ or $-(J_{y}+J_{z})/4$, respectively for $J_y$ or $J_x$ dominant.

%%%%%%%%%%%%%%%%%%%%%%%%%%%%%%%%%%%%%%%%%%%%%%%%%%%%%%%%%%%%
The so-called ring exchange term, which writes $J_{\rm ring} = 3(J_y + J_z)^3/(16J_x^2)$ for a dominant $J_x$, defines a bandwidth of photon excitations with an energy scale of the order of few $\mu$eV for our optimal parameter sets.
As a result, despite the excellent energy resolution, the details of the photon dispersion remain inaccessible, but are nevertheless recorded in the ``quasi-elastic'' signal. Furthermore, these excitations are related to spin components along the $x$ axis of the local coordinate frame, and should manifest in the orbital scattering.

At this point, it is crucial to note that our optimal parameter sets involve a non-zero $J_{xz}$. Using a rotation by an angle $\theta$ defined by $\text{tan}(2\theta)=2J_{xz}/(J_x-J_z)$ about the $y$ axis, $J_{xz}$ can be eliminated from Eq.~\ref{eq:ham} so that the actual relevant variables are “tilted” spins with non-zero projections onto both the initial $z$ and $x$ directions defined in $H_{nn}$ (Eq. \ref{eq:ham}) \cite{DO-PRL,Benton16}. As a result, QSI correlations emerge among the $\tilde{x}$ spins of the rotated frame, and project onto both $z$ and $x$, hence giving rise to both non-zero dipole and orbital neutron cross-sections respectively. We thus expect that a minute $J_{xz}$ leads to a drastic increase in dipolar `quasi-elastic' scattering at the expense of octupolar scattering. 
This naturally explains the simultaneous observation of the scattering displayed in Fig.~\ref{Fig1}d and ~\ref{Fig1}f, which should be understood as the manifestation of the {\it same} ‘photon’ signal along $\tilde{x}$, seen through dipole and orbital neutron cross-sections, respectively.

The $Q$ dependence of the `quasi-elastic' dipolar scattering integrated over $E=[-11,11]~\mu$eV (Fig.~\ref{Fig1}d), peaking around 0.6 \AA$^{-1}$ with a clear drop of intensity at lower $Q$, is a signature of the quantum nature of the ground state, distinct from classical spin ice (CSI) where constant elastic scattering is expected at low $Q$. To illustrate this point, Fig.~\ref{Fig1}d shows fits using analytical models of the scattering for a CSI (black curve) and a QSI (red curve).

\begin{figure}[!b]
\includegraphics[scale=0.32]{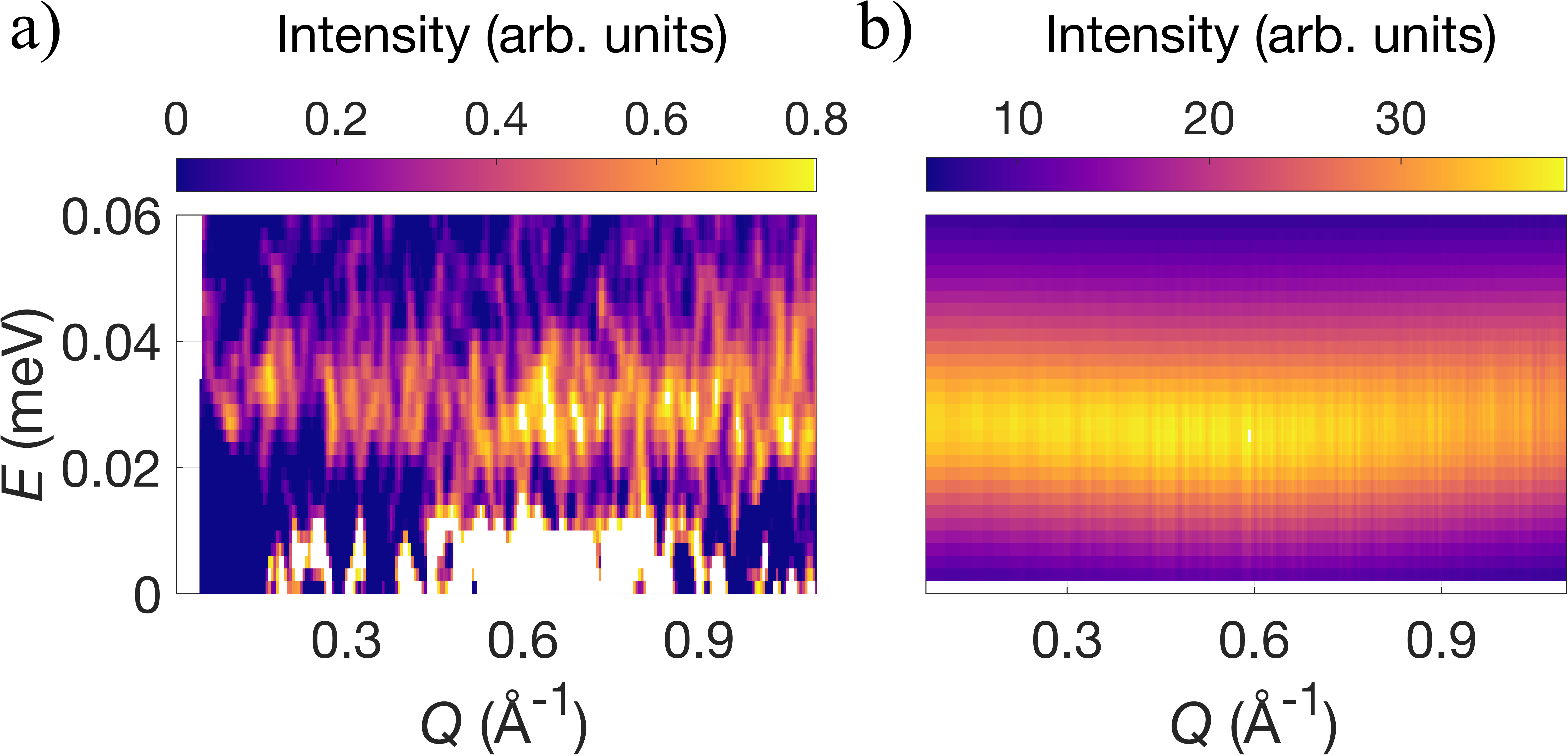}
\centering
\caption{Comparison of the dynamical structure factor as a function of energy and absolute momentum (a) observed (data from Fig.~\ref{Fig1}(c)) and (b) simulated (using parameter set $a$). The latter was obtained from MD simulations for 1024 sites and re-scaled by $\beta E/(1-$exp$(-\beta E))$ ($\beta=1/k_BT$ with $k_B$ the Boltzmann factor and $E$ the neutron energy transfer).}
\label{Fig3}
\end{figure}

To further check consistency with the INS data, one of the optimal parameter sets was used to perform a semi-classical molecular dynamics (MD) simulation ~\cite{ncnf.zhang,CZO_US_2,CZO_CA_2}, computing the energy- and momentum-resolved dynamical structure factor. The resulting spectrum displays a continuum of excitations centered around 0.021 meV, in agreement with the experiment (Fig.~\ref{Fig3}), and interpreted as arising from QSI spinons. The high energy resolution was here crucial to disentangle the dipolar `quasi-elastic' \textit{photon} signal (Fig.~\ref{Fig1}d) from the inelastic \textit{spinon} signal (Fig.~\ref{Fig1}e).

To illustrate the dramatic influence of the $\theta$ angle induced by $J_{xz}$ and independently check our FTLM results, we analyze the data obtained with thermal neutrons shown in Fig.~\ref{Fig1}f over the entire $Q$ range. We introduce a single variable $\phi$ as a continuous parametrization of the doublet, using $\ket{\phi} = \cos\phi \ket{m_J = + 3/2} + \sin\phi \ket{m_J = - 3/2}$ and $\ket{\bar \phi} = -\sin\phi \ket{m_J = + 3/2} + \cos\phi \ket{m_J = - 3/2}$. $\phi=0 and \pi/4$ respectively describe a purely dipolar and octupolar doublet. We assume ice-like correlations, i.e. two-$\ket{\phi}$-two-$\ket{\bar\phi}$ on each tetrahedron \cite{supp}, and calculate the dipole and orbital cross sections using both dipolar and octupolar magnetic form factors. We then estimate $\phi$ from the comparison with the experiment. The optimal value $\phi_{\rm opt} =\pm 0.264~\pi$ indicates a mixed dipolar-octupolar character, yet dominated by octupoles. Deviations from $\phi_{\rm opt}$ significantly alter the amplitudes of dipolar and octupolar ‘photon’ scattering, demonstrating the high sensitivity to the wavefunction’s composition. Translating $\phi_{\rm opt}$ into $\theta$, to ensure comparison with parameters from works on~\cezr~\cite{CZO_CA_2,CZO_US_4} and ~\cesn~\cite{CSO_CA}, gives $\theta_\text{opt}=\pm0.028~\pi$. Remarkably, this independent analysis agrees well with our FTLM results giving \textit{e.g.} $\theta=-0.091\pi$ for parameter set (c) of Eq.~\eqref{eq:optimal-params2}. 
Although the absolute values of the two fittings are different by a factor of $3$, we note that the agreement is actually very good: given the range of $\theta$ is from $0$ to $\pi$ ($\pm\theta$ are indistinguishable), the two independent fittings agree within a window of $(0.028\pi-0.0091\pi)/\pi= 1.9\%$ of the entire possible range of $\theta$. 
Physically, this suggests a  dominant $J_x \gg J_z$, and a very small $J_{xz}$, that is, the spin liquid is a mixture of dipole and octupole moments, with the octupole moments being dominant.

Observing dipolar and octupolar signals in Ce$_2$Hf$_2$O$_7$ informs us directly about the dual nature of the degrees of freedom forming the spin liquid state. Interestingly, a mixed dipolar-octupolar ground state is also proposed in \ndzr~\cite{Xu20, Leger21}. In this particular case, the $\cos \theta$ projection factor naturally explains the strong reduction of the long-range all-in-all-out order parameter. In the present study, however, this dual nature is directly evidenced by the observation of both the dipolar and octupolar signals.

In summary, our analysis suggests that \cehf is a $\pi$-flux QSI characterized by correlations among both dipoles and octupoles. Fine-tuning of $J_{xz}$ is  crucial to make them observable both in the respective dipolar and orbital neutron cross sections. 
Consistent with theoretical expectations, the spinon continuum appears in the dipolar channel above a small gap that could be evidenced thanks to high resolution neutron data. While examples of long-range multipolar ordering have been reported in literature~\cite{Iwahara22,PourovskiiPNAS2021}, our work suggests that \ce pyrochlores constitute a unique example of {\it quantum liquids} involving such correlations of distinct multipoles, opening the way to novel promising studies.

We thank M. Kenzelmann, F. Desrochers and Y.-B. Kim for fruitful discussions.
We acknowledge funding from the Swiss National Science Foundation (project No. 200021\_179150) and European Commission under grant agreement no. 824109 European 'Microkelvin Platform'. This work is also based on experiments performed at the Swiss spallation neutron source SINQ (Paul Scherrer Institute, Switzerland) and at the Institut Laue-Langevin (Grenoble, France, doi:10.5291/ILL-DATA.4-05-813).  J.A.Q. and N.G. acknowledge the support of the Canada First Research Excellence Fund (CFREF) and technical assistance from S. Fortier and M. Lacerte. H.Y. and A.H.N. were supported by the National Science Foundation Division of Materials Research under the Award DMR-1917511. H.Y. is also supported by the 2024 Toyota Riken Scholar Program from the Toyota Physical 
and Chemical Research Institute, and the  Grant-in-Aid for Research Activity Start-up from Japan Society
for the Promotion of Science (Grant No. 24K22856).
A.B. and H.J.C. acknowledge the support of Florida State University and the National High Magnetic Field Laboratory. The National High Magnetic Field Laboratory is supported by the National Science Foundation through NSF/DMR-1644779 and DMR-2128556 and the state of Florida. H.J.C. was also supported by NSF CAREER grant DMR-2046570. We thank the Research Computing Center (RCC) and Planck cluster at Florida State University for computing resources.
\bibliography{CHO_PRL_Biblio}
\end{document}